\documentclass[aps,prx,twocolumn,superscriptaddress,longbibliography]{revtex4-1}
\usepackage{color}
\usepackage{fancyhdr} %页眉页脚页码设置
\usepackage{varwidth} % 可变宽度的 parbox
\usepackage{graphicx,subfigure}  %排版，行间距
\usepackage{bm} %加粗文本
\usepackage{listings}%插入代码字体设置
\usepackage{enumerate}%
\usepackage[colorlinks,citecolor=blue,linkcolor=blue]{hyperref}
\usepackage{amsmath}%公式aligned
\usepackage{threeparttable}%表格加注释
\usepackage{graphicx}
\usepackage{amssymb}
\usepackage{amsthm}
\usepackage{rotating}
\usepackage{multirow}
\def\bl#1{\textcolor{blue}{#1}}

\begin{document}

\title{Atomistic characterization of the SiO$_2$ high-density liquid/low-density liquid interface}

\author{Xin Zhang}
\affiliation{State Key Laboratory of Precision Spectroscopy, School of Physics and Electronic Science, East China Normal University, Shanghai 200241, China}
\author{Brian B. Laird}
\thanks{blaird@ku.edu}
\affiliation{Department of Chemistry, University of Kansas, Lawrence, KS 66045, USA}
\author{Hongtao Liang}
\affiliation{State Key Laboratory of Precision Spectroscopy, School of Physics and Electronic Science, East China Normal University, Shanghai 200241, China}
\author{Wenliang Lu}
\affiliation{School of Science, Changzhou Institute of Technology, Changzhou, Jiangsu 213032, China}
\author{Zhiyong Yu}
\affiliation{State Key Laboratory of Precision Spectroscopy, School of Physics and Electronic Science, East China Normal University, Shanghai 200241, China}
\author{Xiangming Ma}
\affiliation{State Key Laboratory of Precision Spectroscopy, School of Physics and Electronic Science, East China Normal University, Shanghai 200241, China}
\author{Ya Cheng}
\affiliation{State Key Laboratory of Precision Spectroscopy, School of Physics and Electronic Science, East China Normal University, Shanghai 200241, China}
\author{Yang Yang}
\thanks{yyang@phy.ecnu.edu.cn}
\affiliation{State Key Laboratory of Precision Spectroscopy, School of Physics and Electronic Science, East China Normal University, Shanghai 200241, China}
\affiliation{Chongqing Institute of East China Normal University, Chongqing 401120, China}

\begin{abstract}
The equilibrium silica liquid-liquid interface between the high-density liquid (HDL) phase and the low-density liquid (LDL) phase is examined using molecular-dynamics simulation. The structure, thermodynamics, and dynamics within the interfacial region are characterized in detail and compared with previous studies on the liquid-liquid phase transition (LLPT) in bulk silica, as well as traditional crystal-melt interfaces. We find that the silica HDL-LDL interface exhibits a spatial fragile-to-strong transition across the interface. Calculations of dynamics properties reveal three types of dynamical heterogeneity hybridizing within the silica HDL-LDL interface. We also observe that as the interface is traversed from HDL to LDL, the Si/O coordination number ratio jumps to an unexpectedly large value, defining a thin region of the interface where HDL and LDL exhibit significant mixing. In addition, the LLPT phase coexistence is interpreted in the framework of the traditional thermodynamics of alloys and phase equilibria.

%BBL01: I don't understand the previous sentence and I can't figure out what is meant enough to suggest a modification. 
%Yang01: I have deleted the words "which differ significantly from the gelation dynamics". and made change to this sentence.

%BBL02: The previous two sentences need to be rewritten - I'm not sure what their point is.  "This behavior was not seen in previous bulk simulations on systematically searching liquid silica phases. In addition, a new conceptual analogy is made, based on the thermodynamics of alloys and phase equilibria, for interpreting the HDL-LDL interface, which consists of the tetrahedral ordering building blocks."
%Yang02: I have tried to modify this sentence, please double check it.

\end{abstract}
\maketitle

\section{Introduction}
Recently, a considerable literature has emerged around the theme of liquid-liquid phase transitions (LLPT) in pure systems. The origin of this concept can be traced back to the two-local-states model, which was proposed in the 1960s to explain the anomalous behavior of water\cite{Rapoport67}. Later, the application of this theoretical model was gradually expanded to other liquids with a tetrahedral network structure, and the phenomenon of LLPT was found even to exist in non-tetrahedral liquids. So far, strong theoretical, computational and experimental evidence of LLPT has been observed in a variety of systems\cite{Rapoport67,Aasland94,Mishima98,Xu15}, ranging from model systems\cite{Jagla99}, to atomic systems(C, Si, P, Ga, Ce, SiO$_2$, Y$_2$O$_3$, La$_{50}$Al$_{35}$Ni$_{15}$, etc.)\cite{Aasland94,Katayama00,Sastry03,Wu12,Cajahuaringa12,Cadien13,Xu15,Xu16,Chen17}, to molecular liquid-atomic liquid transition systems(H$_2$)\cite{Peter18}, and to molecular systems(H$ _2$O, triphenyl phosphite, etc.)
\cite{Mishima98,Tanaka04,Palmer14,Gallo16,Singh16,Kim20}.

Intensive efforts (mostly theoretical or computational) have been devoted to examining the fundamental nature of this novel structural phase transition \cite{limmer11} and to the location of the LLPT critical point within the metastable realm\cite{Hestand18,Tian19,Debenedetti20}. Recently, Kim, et al. combined X-ray lasers with infrared femtosecond pulses to directly observe LLPT in bulk water at 205K at a pressure between ambient and 3.5 kbar. This experimental study confirms the existence of the LLPT, which has been under debate for over 50 years.\cite{Kim20}.\par

In a natural extension to the previous studies focused on the fundamental bulk phase properties of LLPT, the present work examines the \
bl{dynamical}, thermodynamic and structural properties of the interface between the two coexisting liquid phases. Implementing a careful characterization to gain enough knowledge of the LLPT interface will significantly benefit future potential applications of LLPT. For example,{\em  i}) in the development of the next-generation phase change storage materials, the two liquid phases may be referred as the ``ON'' and ``OFF'' states, and the read/write speeds in data storage\cite{Rao19,Zalden19} is determined by the dynamical properties of the LLPT interface; {\em ii}) the densities of silica glass are determined to a large extent by the densities of the quenching liquids\cite{Mishima85,Handle18,Woutersen18,Bachler19}. A precise characterization and regulation of the possible two liquid density states\cite{Chen17}, under high pressure and temperature conditions (e.g., during ultrafast laser processing with heat accumulations\cite{Cheng13}), could be essential for the potential ultra-fine tuning of refractive index modification\cite{Schaffer03,Hnatovsky06}.

The atomic simulation-based interface characterization methodology has been developed in liquid-vapor interface\cite{Watanabe12}, solid-liquid interface\cite{Davidchack06,Yang12,Liang18}, and the liquid-liquid interface between two immiscible liquids\cite{Zhang95}. The characterized interface properties, on the one hand, can be used as input parameters in mesoscale modeling and in the prediction of microstructure evolution with subsequent comparison to experiment\cite{Wu15,PeterGalenko19}. On the other hand, the thermodynamic properties distributions across the interface are necessary to the formulation of quantitative theories\cite{David82,Curtin86,Wu15,Wu16,Xu20}  for the inhomogeneous fluid system, e.g., the development of the (non)equilibrium Ginzburg-Landau (GL) theory for a pure crystal-melt interface requires the knowledge of the density-wave-related order parameter profiles\cite{Xu20}. Transplanting the well-developed paradigm of interface characterization to the LLPT system is needed so that both the multi-scale modeling of the microstructure evolution and the quantitative thermodynamic/kinetic theory of the LLPT interface systems becomes possible. Unfortunately, to the best of the authors' knowledge, no comprehensive studies of the liquid-liquid interface in LLPT systems have been carried out to date.

The current work is motivated by a recent simulation study\cite{Chen17}, in which Chen et al. identified the stability limits of a tetrahedral model of liquid silica. In addition, they observed that both the LLPT coexistence conditions and structural relaxation in the coexisting low and high-density phases are accessible using molecular-dynamics (MD) simulation. Their MD simulation studies confirmed the existence of a first-order phase transition between the high-density liquid (HDL) and the low-density liquid (LDL) phase of the SiO$_2$. However, only one property (density) along the HDL-LDL interface\cite{Palmer18} was reported. In this work, we perform a comprehensive simulation analysis of the the thermodynamic, structural, and dynamical properties of the SiO$_2$ HDL-LDL interface through the calculation of profiles across the interface of a number of properties including number density, stress, potential energy, coordination number, tetrahedral order parameter, diffusion coefficient, structural relaxation time, and excitation indicator function. Our study provides a new level of understanding of the structure and dynamics of the the LLPT interfaces. The data presented here could be used as critical input parameters for the 
mesoscale modeling and the quantitative theory development for LLPT kinetics.

\section{Methods}
\subsection{Simulation details}

In this work, we use the same model potential for liquid silica used by Chen, et al.\cite{Chen17} in their study of LDL/HDL coexistence; namely a modified version (mWAC) of the potential due to Woodcock, Angell, and Cheeseman\cite{Woodcock76,vanBeest90}, in which the tetrahedral order of liquid silica is enhanced by reducing the electrostatic interactions among ions.  The use of this model to study LLPT had been earlier suggested by Lascaris, et al.\cite{Lascaris16}. In this model, the non-bonded pair potential between ions is expressed as,
\begin{equation}
U_{\mathrm{WAC}}\left(r_{i j}\right)=A_{i j} \exp \left(-B_{i j} r_{i j}\right)+f^{2} \frac{1}{4 \pi \varepsilon_{0}} \frac{z_{i} z_{j} e^{2}}{r_{i j}}
\label{eq1}
\end{equation}
in which, the first term on the right side of Eq.\ref{eq1} represents the short-range ion-ion interactions, $r_{i j}$ is the separation distance between two ions, $A_{\rm SiSi}=1.917991469\times10^{5}$ kJ/mol, $A_{\rm SiO}=1.751644217\times10^{5}$ kJ/mol, $A_{\rm OO}=1.023823519\times10^{5}$ kJ/mol, $B_{ij}=34.48\; {\rm nm}^{-1}$, yielding effective ionic radii of $\sigma_{\rm Si}=0.1301$ nm and $\sigma_{\rm O}=0.1420$ nm\cite{Lascaris14}. The second term is the Coulombic term, in which $z_{\rm Si}=+4$ and $z_{\rm O}=+2$ are original charges on Si and O ions, respectively, $e$ is the elementary charge and $\varepsilon_{0}$ is the vacuum dielectric constant. A scaling factor $f$ was introduced by Lascaris\cite{Lascaris16} to adjust the charge on ions, we employ here the value suggested by Chen, et al.\cite{Chen17} of $f=0.84$, so that a fully realized LLPT is computationally accessible. Similar to the behavior seen in the analogous LLPT in the ST2 water model\cite{Guo18b}, Palmer, et al.\cite{Palmer18} have demonstrated for the mWAC model that spontaneous liquid-liquid phase separation is insensitive to system size over a range of particle number spanning more than two orders of magnitude (a few thousands to half million), thus no strong finite-size effect is expected in this work.

The MD simulations are performed using open source software LAMMPS released by the Sandia National Lab\cite{Steve95}. All simulations are performed with a constant time step of 1 fs, using a Nos\'{e}-Hoover thermostat\cite{Hoover85} (thermostat relaxation time 0.1 ps) to maintain a temperatures of 3100K. In addition, periodic boundary conditions are applied along three orthogonal directions. In some of the simulations, an Anderson barostat (relaxation time 1.0 ps) is employed to regulate pressure along the direction normal to the interface. The long-range Coulombic interactions are calculated using the the particle-particle-particle-mesh (PPPM) solver along with a real-space cut-off of 10\r{A} and a relative root-mean-square force error of approximately 10$^{-4}$.

The construction of the equilibrium HDL-LDL interface follows the procedures described in Ref.\onlinecite{Yang12}. The initial configurations are constructed by first creating separate HDL and LDL samples. A sample of charge neutral bulk LDL containing 4000 oxygen atoms and 2000 silicon atoms is prepared first, with $Np_zA_{xy}T$ MD simulations ($T=3100$ K, $p=0.4$ GPa, mass density 1.55 g/cm$^{-3}$). The cross-section for this LDL sample is chosen to be square, with the fixed dimensions $L_x$ and $L_y$ in the $x$ and $y$ directions, respectively and a cross-sectional area $A_{xy}=L_{x}\times L_{y}=30$\AA$\times$30\AA. A separate sample of HDL containing 6000 atoms is created, with $Np_zA_{xy}T$ MD simulations ($T=3100$ K, $p=0.4$ GPa, mass density 2.10 g/cm$^{-3}$), using the same cross-sectional dimensions as the corresponding LDL sample. The initial interface configuration is assembled by conjoining the LDL and the HDL samples at their common cross-section and applying periodic boundary conditions to the conglomerate. A separation distance of around 2\r{A} between the HDL and the LDL samples is chosen to avoid artificial atomic position overlapping during the concatenation of the two samples. The assembled HDL-LDL interface is then equilibrated using $Np_zA_{xy}T$ MD simulations under $T=3100$ K and $p=0.4$ GPa. Note that, as suggested by Chen et al.\cite{Chen17}, at a few temperatures below 3350 K at pressures around 0.4 GPa, the density-dependent free energy functions exhibit double-basin structures, e.g., for the systems (3000 K, 0.37 GPa) and (3150 K, 0.5 GPa), the HDL phase (mass density $\approx$ 2.10 g/cm$^{-3}$) and the LDL phase (mass density $\approx$ 1.55 g/cm$^{-3}$) are predicted to coexist with large free energy barriers (3.8 and 6.5 $k_\mathrm{B}T$, respectively).

We follow the initial $Np_zAT$ run with a long $NVT$ MD simulation, lasting at least 300 ns, in order calculate the equilibrium averages and interfacial profiles. The temperature and pressure proﬁles are examined throughout the MD simulations to ensure that no temperature gradients are present and the pressure component $p_z$ is constant at 0.4 GPa in the equilibrium simulations used for the characterization analysis. We also monitor the excess stress in the bulk HDL and LDL regions to ensure that the hydrostatic condition applies in the bulk phases. To ensure that we use the proper value of $L_z$ in starting the $NVT$ simulations we calculate the average $L_z$ from the final 150 ns of the $Np_zAT$ run and initiate the $NVT$ run from an $NP_zAT$ configuration with the instantaneous $L_z$ that matches the average. The final system dimensions during the $NVT$ run are $L_{z}=233.4$\r{A}, $L_{x}=L_{y}=30.1$\r{A}, with an aspect ratio $L_x:L_y:L_z = 1:1:7.8$ and mean mass density converged to $1.87 {\rm g/cm^{-3}}$, which has been suggested for formation of a stable planar HDL-LDL interface\cite{Chen17,Palmer18,Singh19}.

%BBL03: There is some confusion here - the temperature is given as both 3100K and 3150K. You also need a couple of sentences discussing why you went to a higher temperature than predicted by Chen, et al.  What was the final simulation temperature for the profiles. The paper does need some discussion as to why the temperature is different than in the Chen paper.

%Yang03: I have modified this part, to clear the confusion. I have asked Xin to construct the equilibrium interface from two bulk samples created at  $T=3100$ K, $p=0.4$ GPa, and the equilibrium interface can be easily obtained, in comparing with previous construction method. Now, throughout the paper, the working temperature are all at 3100K. and I have added a few sentences, following your suggestion, in the above paragraph, please have a check.

\subsection{Characterization details}
The interface is characterized primarily through the calculation of profiles that measure changes in a given quantity as the interface is traversed along the direction normal to the interfacial plane (here defined as the $z$ axis). The profiles are determined by binning (using either fine-grained or coarse-grained bins) the $z$ direction and averaging the quantities of interest within each bin over the $xy$ plane. In this work, three categories of (thermodynamic, structural, and dynamic) properties are characterized.

\subsubsection{Thermodynamic properties}
The thermodynamic properties investigated in this work include number density, stress and potential energy. Fine-grained profiles of these three properties are calculated using the protocols outlined below. The scale of the fine-grain (or the bin size mentioned below) is chosen as $\delta_z=0.05$ \AA, the results is attained by averaging over 200 million trajectories.\\

\textbf{\emph{Density profile}:} The number density profile $\rho$($z$) is defined by
\begin{equation} \label{rho}
\rho(z)=\frac{\left\langle N_{z}\right\rangle}{A_{x y} \delta_z}
\end{equation}
where $\delta_z$ is the bin spacing, ${\left\langle N_{z}\right\rangle}$ is the average number of atoms (we do not distinguish Si and O) in the bin defined by $z-\delta_z/2<z<z+\delta_z/2$. Because of periodic boundary conditions are applied here, there are two liquid-liquid interfaces in the simulation box. The density profile (and the profiles mentioned below) averages the information of two interfaces in the simulation box to improve the statistics.\\

\textbf{\emph{Stress profile}:} The stress profile $S(z)$ is defined as the difference between normal and transverse pressure components. Its spatial distribution is determined as
\begin{equation} \label{S}
S(z)=p_{z z}(z)-\frac{1}{2}\left[p_{x x}(z)+p_{yy}(z)\right].
\end{equation}

The fine-grained microscopic pressure components are determined by binning the $z$ axis (with bin size $\delta_z$) and calculating the sum of the negative per-atom stress tensors $s_{i}^{\alpha \beta}$\cite{Thompson09}, divided by the bin volume:
\begin{equation} \label{P}
P_{\alpha \beta}(z)=-\frac{\left\langle\sum_{i}^{N_{z}} s_{i}^{\alpha \beta}(z)\right\rangle}{A_{xy} \delta_z},
\end{equation}
where the summation runs over $N_{z}$ atoms located in the interval $z-\delta_z/2<z<z+\delta_z/2$.
\begin{equation} \label{s}
s_{i}^{\alpha \beta}=-\left[m v_{i \alpha} v_{i \beta}+\frac{1}{2} \sum_{j=1}^{N_{n}}\left(r_{i_{a}} F_{i_{\beta}}+r_{j_{a}} F_{j_{\beta}}\right)\right],
\end{equation}
in which, $m$ is the atom mass, $v_{i}$ is the velocity of atom $i$. The indices ${\alpha}$ and ${\beta}$ can be $x$, $y$, or $z$. $r_{ij}$ and $F_{ij}$ are the distance and force between atom $i$ and atom $j$ connected with a pairwise potential. $N_{n}$ is the number of atoms to atom $i’s$ neighboring atoms. For a regular liquid-liquid interface under hydrostatic pressure conditions, $S(z)$ should be zero away from the interfacial region\cite{Zhang95,Andrij99,Wang11,Palafox11,Yang12,Wen17}. As we will see below, $S(z)$ in the current simulations satisfies this condition.\\

\textbf{\emph{Potential energy profile}:} The potential energy profile, $\rho_e(z)$, is computed by averaging the potential energy ${\left\langle U_{z}\right\rangle}$ within each bin and dividing by  $\left\langle N_{z}\right\rangle$ of the bin:
\begin{equation} \label{U}
\rho_e(z)=\frac{\left\langle U_{z}\right\rangle}{\left\langle N_{z}\right\rangle}.
\end{equation}

\subsubsection{Structural properties}
The structural properties investigated in this work are characterized by the tetrahedral order parameter and the coordination numbers. Coarse-grained profiles of these two structural properties are calculated as the following way. The scale of the coarse-grain bin size is chosen to be $\Delta z=4.0$ \AA,  which corresponds to the location of the first minimum in the Si-Si (or O-O) radial distribution function, $r_{\rm min}$. We divide a total of 4000 $NVT$ trajectories into 10 blocks, each block contains two independent HDL-LDL interfaces, giving a total of 20 samples for the block averaging and determining the statistical uncertainty.\\

\textbf{\emph{Coordination number profiles}:}
The coordination number $n_{\rm O}(z)$ is the average number of O ions surrounding one Si ion, defined by
\begin{equation}
n_{\mathrm{O}} (z)= 4 \pi \rho^{z}_{\mathrm{O}} \int_{0}^{r_{\mathrm{min}}} r^{2} g^{z}_{\mathrm{SiO}}(r) \mathrm{d}r,
\end{equation}
$\rho^{z}_{\rm O}$ and $g^{z}_{\rm SiO}$ are the number density of the O ions and the Si-O radial distribution function in $z-\Delta z/2<z<z+\Delta z$, respectively. Similarly, the coordination number $n_{\rm Si}(z)$ is defined as,
\begin{equation}
n_{\mathrm{Si}} (z)= 4 \pi \rho^{z}_{\mathrm{Si}} \int_{0}^{r_{\mathrm{min}}} r^{2} g^{z}_{\mathrm{SiO}}(r) \mathrm{d}r.
\end{equation}
\\

\textbf{\emph{Tetrahedral order parameter profile}:} We use per-atom tetrahedral order parameters $q_{i}$\cite{Debenedetti01} of Si ions, as one of the structural descriptors, as a way of quantifying the liquid local structure along $z$, 
\begin{equation}
q_{i}=1-\frac{3}{8} \sum_{j=1}^{3} \sum_{k=j+1}^{4}\left(\cos \Theta_{i j k}+\frac{1}{3}\right)^{2} ,
\end{equation}
where $\Theta_{ijk}$ denotes the angle between the silicon atom $i$ and two of its four nearest silicon atoms $j$ and $k$. The $q_{i}$ values range from -3 to 1, with  $q_{i}$=1 representing a perfect local tetrahedral structure. The tetrahedral order parameter $Q$ of the coarse-grain bin is defined as the weighted average over the $q_{i}$ of all silicon ions within one specific coarse-grain bin,
\begin{equation}
Q(z)=\int_{-3}^{1} f\left(q_{i},z\right) q_{i} \mathrm{d} q_{i},
\end{equation}
where $f(q_{i},z)$ is the normalized probability distribution function, in  $z-\Delta z/2<z<z+\Delta z$. When $Q$ = 1, all atoms are arranged in a perfect tetrahedral structure, while $Q$ = 0 corresponds to a completely random isotropic arrangement as in an ideal gas.
\subsubsection{Dynamical properties}
The dynamical properties investigated in this work include the diffusion coefficient, the structural relaxation time, and an excitation indicator function. The scale of the coarse-graining is also chosen as $\Delta z=4.0$ \AA.\\

\textbf{\emph{Diffusion coefficient profile}:} All atoms at an initial time $t_{0}$ are allocated into the corresponding coarse-grained bins; we then trace their diffusive dynamics within a time window $t_{D}$. The diffusion coefficient in each bin is then determined from calculating the average mean-square displacement (MSD) of the ions,$\left\langle\left[\mathbf{r}_{j}(t)-\mathbf{r}_{j}(0)\right]^{2}\right\rangle_{z}$, from extracting the slope of the linear dependence regime in the MSD versus time,
\begin{equation}
D(z)=\lim _{t>t_{D}} \frac{1}{6} \frac{d}{d t}\left\langle\left[\mathbf{r}_{j}(t)-\mathbf{r}_{j}(0)\right]^{2}\right\rangle_{z}.
\label{diff}
\end{equation}
The average MSD in each coarse-grained bin is calculated from averaging over 100 independent time origins separated by 10 ps each over more than ten replica $NVT$ runs. Note that, $t_{D}$ is smaller than the average time required for an atom diffuse halfway across one bin.\\

\textbf{\emph{Structural relaxation time profile}:} To determine the structural relaxation time profile, one should calculate the incoherent intermediate scattering functions (ISF) over each coarse-grained bins first,
\begin{equation} \label{F}
F^{z}_{s}(k_\mathrm{Si,O}, t)=\frac{1}{N_{z}} \sum_{j=1}^{N_{z}}\left\langle\exp \left\{i k_\mathrm{Si,O} \cdot\left|\mathbf{r}_{j}(t)-\mathbf{r}_{j}(0)\right|\right\}\right\rangle_{z},
\end{equation}
in which, $N_z$ is the number of Si or O ions in $z-\Delta z/2<z<z+\Delta z$. The wave number $k_\mathrm{Si,O}=2\pi/d_\mathrm{Si,O}$ corresponding to wavelengths equal to the average nearest neighbor distance between silicon ions or oxygen ions in bulk liquid silica, respectively, $d_\mathrm{Si}=3.465$\r{A}, $d_\mathrm{O}=2.815$\r{A}. The structural relaxation time $\tau_\mathrm{Si}$ and $\tau_\mathrm{O}$ in each coarse-grained bin are defined as the times after which the $F^{z}_{s}(k_\mathrm{Si}, t)$ and $F^{z}_{s}(k_\mathrm{O}, t)$ have decayed from 1 to $1/e$.

\textbf{\emph{Excitation indicator function profile}:} Keys et al.\cite{Chandler11} introduced excitation indicator function to identify the atom excitation for a glass-forming liquid. Here we use this indicator function to identify the dynamical inhomogeneity between HDL and LDL. Following Keys et al., one atom is associated with an excitation that persists a displacement greater than $a$ over a duration time $\Delta t$, these excitations are identified by computing for every trajectory the functional over at least $\Delta t$, 
\begin{equation} \label{h}
h_{i}\left(t, t_{a} ; a\right)=\prod_{t^{\prime}=t_{a} / 2-\Delta t}^{t_{a} / 2} \theta\left(\left|\overline{\mathbf{r}}_{i}\left(t+t^{\prime}\right)- \overline{\mathbf{r}}_{i}\left(t-t^{\prime}\right)\right |-a\right), 
\end{equation}
here, $\theta(x)$ is the Heaviside step function $\theta(x)=1$ or 0 for $x\geq0$ or $<0$, respectively. The products are over a trajectory (consists of each 1 fs time-step) that extends for a time $t_{a} > \Delta t$. $t_{a}$ is a plateau or commitment time, $\Delta t$ is a instanton time, which is the shortest time separating the initial and final sojourns, and $a$ is the displacement length. $h_{i}\left(t, t_{a} ; a\right)=1$ if the atom is associated with an excitation at time $t$, and $h_{i}\left(t, t_{a} ; a\right)=0$ otherwise. We take $\tau$ (for bulk HDL) calculated by Eq.(\ref{F}) as the value of $t_{a}$, and choose half of an atom diameter approximately as the value of $a$\cite{Chandler78,Peter90,Chandler11}. Specifically, ${\Delta t}=500$ fs, $t_a=2000$ fs, $a=0.68$ \AA.

\subsubsection{Gibbs Dividing Surface}
In the framework of the Gibbs surface thermodynamics, a critical concept for quantifying physical properties of an interface is the Gibbs Dividing Surface(GDS)\cite{William91}. In this work, this imaginary surface separating the two phases is chosen as the position where the excess number of atoms (regardless Si or O) is zero, 
\begin{equation}
2\Gamma=\frac{N}{A_{x y}}-\rho_\mathrm{LDL} L_{z}-\left(\rho_\mathrm{HDL}-\rho_\mathrm{LDL}\right) L_\mathrm{HLD}=0,
\end{equation}
in which, the factor of 2 is present because there are two independent interfaces in the simulation, $\Gamma$ is the excess number of atoms. $\rho_\mathrm{HDL}$ and $\rho_\mathrm{LDL}$ are the number densities in the bulk HDL and LDL, respectively. $L_\mathrm{HDL}$ and $L_\mathrm{LDL}$ are the corresponding $z$-direction lengths of the two liquid phases. In all the profiles shown below, $z=0$ represents the location of the GDS, $z<0$ for HDL and $z>0$ for LDL.\\

\section{Results and discussion}

\subsection{Thermodynamic properties}

\begin{figure}[!htb]
	\centering
	\includegraphics[width=0.5\textwidth]{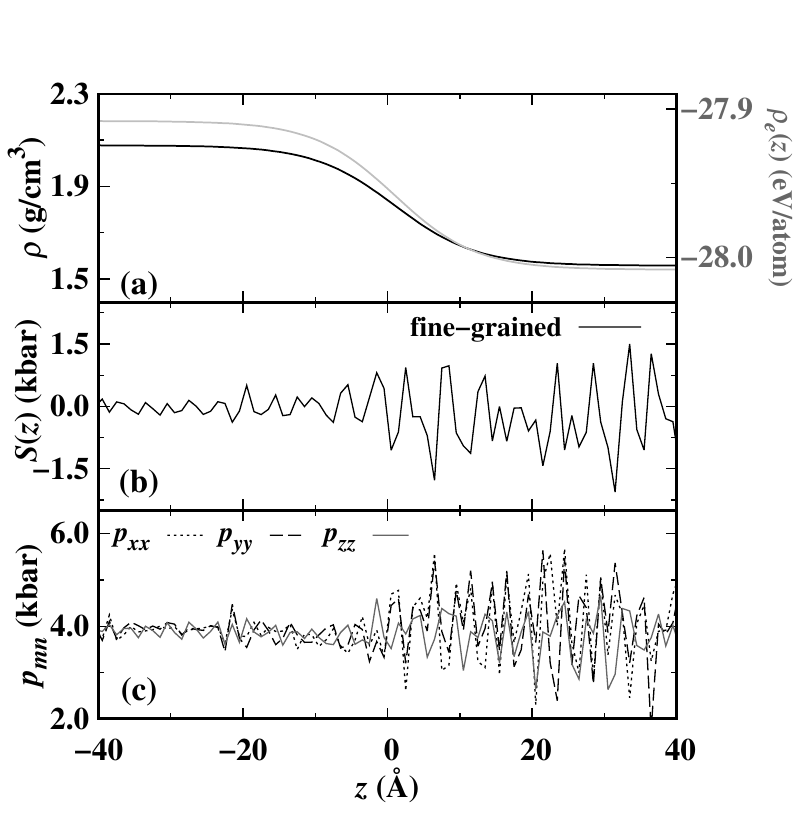}
\caption{(a) Fine-grained density (black) and energy (gray) profiles. (b) Fine-grained stress (solid line) and the corresponding smoothed coarse-scale stress profile (dashed line) using the finite impulse response filter\cite{Davidchack98}. (c) Fine-grained pressure components profiles. For the silica HDL-LDL interface under $T=3100$ K, $p=0.4$ GPa. $z=0$ correspond to GDS, with $z<0$ for HDL and $z>0$ for LDL.\label{1}}
\end{figure}

A closeup of fine-grained profile of the number density is shown in FIG.\ref{1}(a). The equilibrium values of the bulk HDL and LDL are calculated from averaging approximately 1/3 of the plateau regions (i.e., $-$60\AA \ $<z<$\ $-$30\AA \ and 30\AA \ $<z<$\ 60\AA) in the two-phase coexistence $\rho(z)$ profile, $\rho_\mathrm{HDL}= 2.08$g/cm$^3$, $\rho_\mathrm{LDL}=1.56$g/cm$^3$ consistent with previous individual bulk liquid simulations\cite{Chen17} under the same temperature and pressure conditions. 
% BBL04: I'm not sure what "1/3 portion" means here - can you define it more explicitly.  
% YY04:   I have defined this region for the average.

The 10-90 width,  $\delta_{10-90}$, of the fine-grained density profile\cite{Davidchack98} is about 8.0 times oxygen atom diameter, $ \sim$23 \AA. (Note: the 10-90 width of an monotonic interfacial profile is defined as the distance over which the quantity of interest changes from 10\% to 90\% of its value in one phase (e.g. LDL)  relative to its value in the other (e.g. HDL) as one traverses the interface from one bulk phase to another.) According to the capillary wave theory, the width of the liquid-liquid interface depends upon both the interfacial free energy and the interfacial cross-section area due to capillary fluctuations\cite{Rowlinson02}. For similar cross sectional areas, the 10-90 width for the current HDL-LDL liquid-liquid interface is more than three times broader than a previous reported liquid-liquid Al-Pb interface\cite{Yang14} at melting temperature of Al.
%BBL05: I'm not certain of this comparison. If I remember correctly, the width of the interface is an intrisic width plus a term that depends on \gamma and area; therefore, one cannot make conclusions about the relative values of \gamma - for example, they could have the same \gamma, but different intrinsic widths. 
%YY05: Yes, you are right Brian, I have removed the words relating to the interfacial free energies -- "suggesting that the interfacial free energy (or interfacial tension) of the current silica HDL-LDL is significantly smaller than that of the liquid-liquid Al-Pb interface, 253 mN/m."

In FIG.\ref{1}(a), Complementing the information provided by the fine-grained density profile is the potential-energy profile (gray line). Both profiles flatten out far away from the GDS and correspond to the bulk regimes of the HDL and LDL. The calculated potential energies in bulk HDL and LDL are -27.9 eV/atom and -28.0 eV/atom, respectively. Each atom in the LDL phase has a deeper potential minimum than in the HDL phase. Both the 10-90 width and the midpoint of the $\rho_e(z)$ profile are consistent with those values for the $\rho(z)$ profile, see in Table~\ref{tab1}.

%~~~~~~~~~~~~~~~
\begin{table}[!ht]
	\caption{The 10-90 width, $\delta_{10-90}$, and the (midpoint) position  relative to the GDS, $z_\mathrm{mid}$, of the different SiO$_2$ HDL-LDL interfacial profiles. Including the  fine-grained density $\rho(z)$ and energy $\rho_e(z)$ profiles, coarse-grained profiles for coordination numbers ($n_\mathrm{Si}(z)$, $n_\mathrm{O}(z)$), averaged tetrahedral order parameter $Q(z)$, diffusion coefficient D(z), and structure relaxation times ($\tau_\mathrm{Si}(z)$, $\tau_\mathrm{O}(z)$).}
\begin{ruledtabular}
	\begin{tabular}{ccccccccc}
		{}&$\rho(z)$&$\rho_e(z)$&$n_\mathrm{Si}(z)$&$n_\mathrm{O}(z)$&$Q(z)$&$D(z)$&$\tau_\mathrm{Si}(z)$&$\tau_\mathrm{O}(z)$\\
		\hline
		$\delta_{10-90}$ &22.7&22.7&31.5&27.9&23.8&33.1&18.7&11.1\\
		$z_\mathrm{mid}$&0.8  & 0.8  &-0.5 &0.0 &2.0 &-4.6 &8.0&7.6\\
	\end{tabular}
\end{ruledtabular}
\label{tab1}
\end{table}

%~~~~~~~~~~~~~~~

The stress profile $S(z)$ is shown in FIG.\ref{1}(b). Both the bulk HDL and LDL are under hydrostatic stress with zero stress, which implies that the $S(z)$ should approach zero far away from the interface on both the HDL and LDL side. The individual pressure component profiles are shown in FIG.\ref{1}(c). The normal $p_{zz}$ and transverse pressure $p_{xx}$ and $p_{yy}$ components are identical, around 4 kbar, in HDL($p_{xx}=3.9\pm0.1$kbar, $p_{zz}=3.9\pm0.1$kbar, $p_{zz}=3.9\pm0.1$kbar) and in LDL ($p_{xx}=4.1\pm0.4$kbar, $p_{zz}=4.1\pm0.4$kbar, $p_{zz}=3.8\pm0.2$kbar). In LDL, large static oscillations in $S(z)$ are seen, in contrast to the relatively smooth shape seen in HDL. The peak values and the periods in these oscillatory structures are pretty robust over hundreds of nanosecond, and they do not show any similarities, compared with those observed in the fine-grained stress profiles of crystalline solids. These local stress features are likely due to the tetrahedral network structure and the long structural relaxation times.

The excess stress $\sigma_\mathrm{ex}$ is given by the integral over $z$ of the stress profile $S(z)$ and, for an equilibrium liquid-liquid interface, is equal to the interfacial free energy $\gamma$.\cite{Kirkwood49} For our current simulations at $3100$K, we integrate the data in FIG.\ref{1}(b) to obtain 
$\sigma_\mathrm{ex} = -6\pm54$ mJ/m$^2$, consistent with a very small value of $\gamma$. Note that $T = 3100$K is just below the estimated critical temperature of 3350K for this system.\cite{Chen17} For a first-order phase transition between two fluid phases, the interfacial free energy vanishes as the critical point is approached, so data indicating a small value of the excess stress is consistent with the proximity of the transition critical point.

\subsection{Structural properties}\par

\begin{figure}[!htb]
	\centering
	\includegraphics[width=0.48\textwidth]{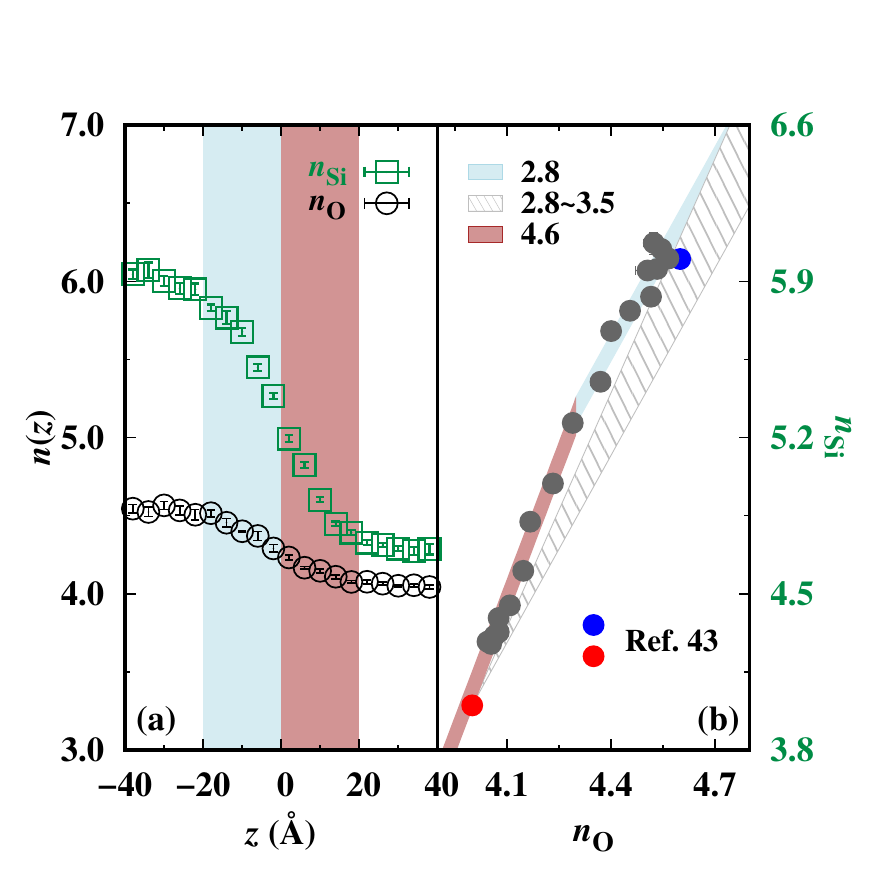}
\caption{(a) Coordination number profile for $n_{\rm Si}$ (smaller circle) and $n_{\rm O}$(larger circle). The error bars represent 95$\%$ confidence levels. (b) The gray shaded area represents the slope interval from $\sim$2.8 to $\sim$3.5 suggested by Lascaris\cite{Lascaris16} on the dependence relationship between $n_{\rm O}$ and $n_{\rm Si}$. The red and blue solid circle are the bulk HDL and LDL data. The light-blue bar and the brown bar are the linear fits to the ($n_{\rm O}$;$n_{\rm Si}$) datasets across the HDL-LDL interface, their slopes are $\sim$2.8 and $\sim$4.6, respectively.\label{2}}
\end{figure}

To obtain a more complete knowledge of the $\rm SiO_{2}$ HDL-LDL interface, we \bl{analyze} the microscopic structural properties (coordination number and tetrahedral order parameter) as functions of distance from the interfacial plane. FIG.\ref{2}(a) shows the coordination number profile. From this plot, one can see that in HDL region, $n_{\rm O}$ and $n_{\rm Si}$ converges to 4.5 and 6.1, respectively, indicating that the Si and O atoms in the HDL are significantly overcoordinated relative to a tetrahedral liquid -  each Si ion in HDL is surrounded by 0.5 additional O ions and 2.1 additional Si ions, over a tetrahedral reference. Hereafter, we employ the notation ($n_{\rm O}$;$n_{\rm Si}$) for the coordination number dataset. In LDL region, one gets ($n_{\rm O}$;$n_{\rm Si}$)=(4.0;4.2), suggesting that the SiO$_{2}$ LDL, under the current $T-p$ condition, exhibits tetrahedral order. Note that, similar results for the bulk HDL(4.6;6.0) and LDL(4.0;4.0) SiO$_2$ under slightly different $T$ and $p$ (3240K and 0.3GPa) have been reported by Lascaris.\cite{Lascaris16}

\begin{figure}[!htb]
	\centering
	\includegraphics[width=0.5\textwidth]{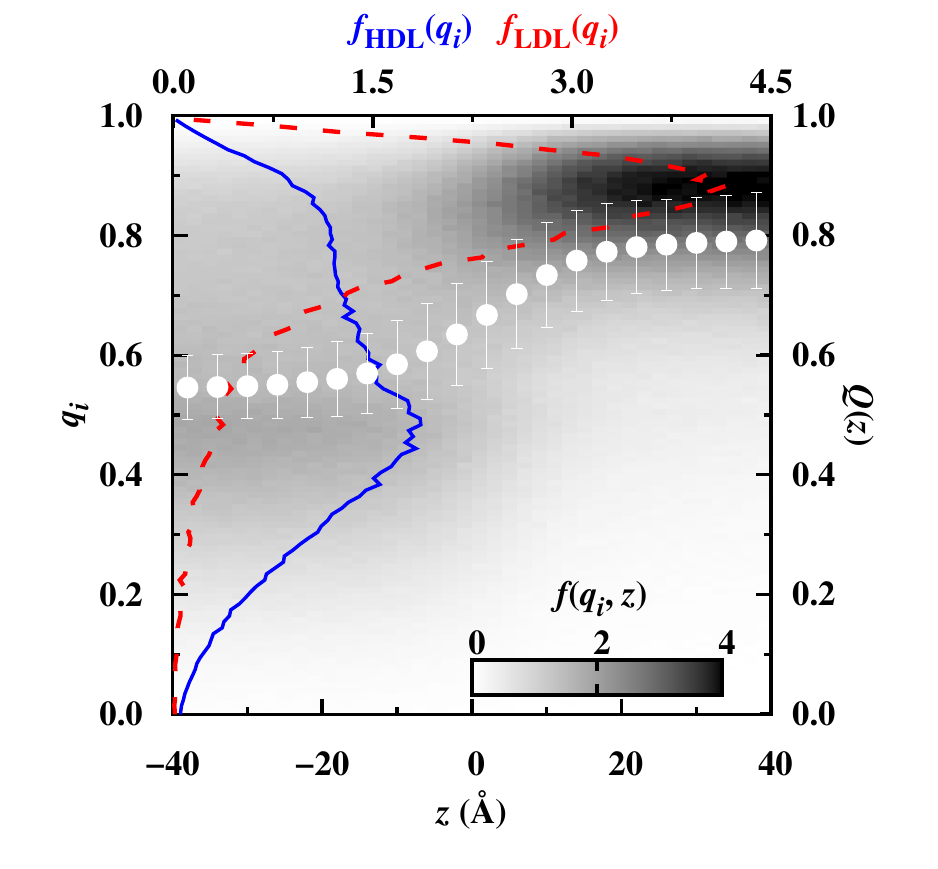}
	\caption{Grayscale contour plot of the probability distribution function of the tetrahedral order parameter across the SiO$_2$ HDL-LDL interfaces, $f(q_{i},z)$. The corresponding coarse-grained profile of the averaged tetrahedral order parameter, $Q(z)$, represented with the white solid circles, is plotted on the top of the $f(q_{i},z)$. In addition, the probability distributions of the atomic tetrahedral order parameter for both the bulk HDL ($f_\mathrm{HDL}(q_{i})$) and the bulk LDL ($f_\mathrm{HDL}(q_{i})$) are plotted with the blue solid line and the red dashed line, respectively.\label{3}}
\end{figure}

From analyzing SiO$_2$ systems over a wide range of density and ion charge magnitudes, Lascaris\cite{Lascaris16} proposed a linear interrelationship between $n_{\rm O}$ and $n_{\rm Si}$, and suggested that LLPT critical point occurs only the $n_{\rm Si}$/$n_{\rm O}$ ratio falls in the value range between $\sim$2.8 to $\sim$3.5. Below $n_{\rm Si}/n_{\rm O}\sim2.8$, the liquid-vapor spinodal dominates over the LLPT critical point, preventing the formation of the metastable LDL; Above $n_{\rm Si}/n_{\rm O}\sim3.5$, the silica liquid remains homogeneous and the LLPT is forbidden.

FIG.\ref{2}(b) plots such correlation between the coordination numbers $n_{\rm Si}$ and $n_{\rm O}$, within the HDL-LDL interface range of $\pm40$ \AA \  shown in FIG.\ref{2}(a). Despite the fact that the coordination numbers ($n_{\rm O}$; $n_{\rm Si}$) for the bulk HDL ($<-20$ \AA) and bulk LDL ($>20$ \AA) agree with Lascaris, a novel $n_{\rm O}$ and $n_{\rm Si}$ correlation for the interface SiO$_2$ is observed. For the interface SiO$_2$ in 0\AA  \ $<z<20$\AA, the $n_{\rm Si}$/$n_{\rm O}$ slope reaches a value of 4.6, which is nearly triple times of the upper limit slope reported by Lascaris. As the HDL-LDL interface is continuously traversed (from LDL to HDL), the $n_{\rm Si}$/$n_{\rm O}$ slope changes from 4.6 to 2.8. Specifically, in region 0\AA \ $<z<20$\AA \ $n_{\rm Si}$ varies its value 4.6 times faster than $n_{\rm O}$; in region $-$20\AA \ $<z<0$\AA, $n_{\rm Si}$ varies its value 2.8 times faster than $n_{\rm O}$. Only the variation in the latter region agrees with Lascaris’s arguments. The significant slope and the subsequent transition within the HDL-LDL interface, could be interpreted from the $n_{\rm O}(z)$ and $n_{\rm Si}(z)$ profiles have different spatial relaxation lengths, i.e, the 10-90 widths of the $n_{\rm Si}(z)$ and $n_{\rm O}(z)$ profiles are about 31.5\AA \ and 27.9\AA, respectively, and the (midpoint) position of the two coordination number profiles separated from each other by a distance of 0.5\AA.

\begin{figure*}[!htb]
\centering
\includegraphics[width=0.85\textwidth]{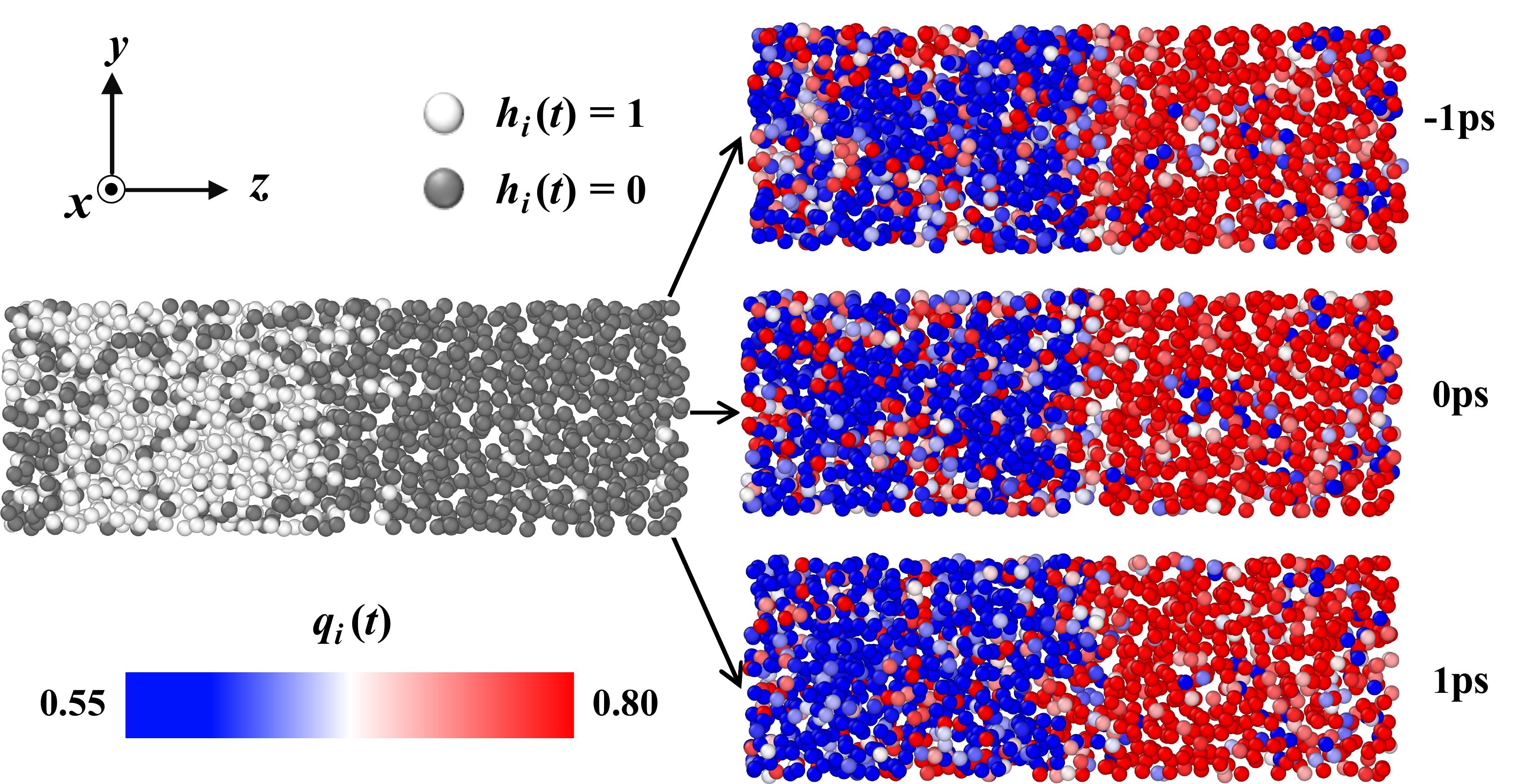}
\caption{Left side, a MD snapshot of the silica HDL-LDL interface, with only Si ions shown, each atom is color coded based on the calculated excitation indicator function. The white color is associated with excitations with a minimum displacement $a$ over a time duration of around 1ps, while the gray color indicates atoms that are relatively immobile. The three successive trajectories (separated by time interval of 1ps) on the right side are color coded based on the instantaneous atomic tetrahedral order parameter $q_i(t)$, the red color stands for the stronger tetrahedral order.\label{8}}
\end{figure*}

It is known that the local thermodynamic properties within interfaces could differ from their corresponding bulk phases (e.g., non-hydrostatic mechanical stress built-up in the liquid-vapor\cite{Moore15} or crystal-melt\cite{Lu21} interfaces; abnormal mutual miscibilities found in the heterogeneous solid-liquid interface\cite{Palafox11}). The $n_{\rm Si}$/$n_{\rm O}$ ratio has been related to the mixing Gibbs free energy $\Delta G_{\rm mix}$ and mixing entropy $\Delta S_{\rm mix}$\cite{Lascaris16} for the mixture of the LDL and HDL phase. A larger $n_{\rm Si}$/$n_{\rm O}$ slope would probably lead to an increase of $\Delta S_{\rm mix}$, or smaller $\Delta G_{\rm mix}$. Therefore, the HDL-LDL interface region with the higher value of the $n_{\rm Si}$/$n_{\rm O}$ ratio indicates a thin layer of interface SiO$_2$ (about 20 \AA) with stronger mixing ability, comparing with the bulk phases, whereas, the rest part of the interfacial liquid SiO$_2$ in the region ($-$20\AA  \ $<z<0$\AA) has regular mixing ability.

%BBL07: Not sure I understand the previous sentence. Please reword.
%Yang07: I have delete this sentence. "We find it pretty interesting is because of this thin layer of interfacial SiO$_2$ is sandwiched between the bulk LDL and the rest portion of the HDL-LDL interface next to the bulk HDL.". And adjusted the sequence of the rest sentences.

We employ the tetrahedral order parameter to reveal the degree of local structural order within the first coordination shell. Overall, the coarse-grained averaged tetrahedral order parameter profile, $Q(z)$ (white solid circles in FIG.\ref{3}) suggests that the degree of local order changes considerably with the spatial variation, LDL has higher tetrahedral order than HDL. We note that the $Q(z)$ profile is shifted towards LDL relative to the $\rho(z)$ and the GDS, indicating that the density of the system relaxes before the tetrahedral order, from HDL to LDL. In addition, the $Q(z)$ profile has a slightly larger 10-90 width (23.8\AA) than the density, energy, and the coordination number profiles.

The $f(q_{i})$ plot reveals the probability distribution of atomic $q_{i}$, for the bulk region of LDL, the $f_{\rm LDL}(q_i)$ distribution has one major peak at $q_{i}$ $\approx$ 0.88, suggesting LDL is characterized by strong tetrahedral order. The peak value is close to that observed in supercooled water, in which a similar degree of the tetrahedral structural order is found\cite{Klameth13}. In addition, the $f_{\rm LDL}(q_i)$ distribution is also consistent with that found in liquid silica a strong glass former) at $T=2300$ K and ambient pressure\cite{Julian16}. The contour plot of the $f(q_i,z)$ in FIG.\ref{3} shows a striking transition from the highly concentrated unimodal distribution of LDL to the asymmetric single peak distribution with a shoulder seen in HDL. This transition is also reflected in the instantaneous snapshots in which atoms are color-labeled based on the magnitudes of $q_i$ in FIG.\ref{8}, a stronger heterogeneity in color is found in the HDL comparing with LDL. The dominant peak in $f_{\rm LDL}(q_i)$ transitions to a shoulder in the $f_{\rm HDL}(q_i)$ distribution, meanwhile, a new peak at $q_i$ $\approx$ 0.43 develops in $f_{\rm HDL}(q_i)$. The local structure is significantly less ordered, because the partition for the low-values of $q_i$ increases while the partition for the high-values of $q_{i}$ (with strong tetrahedral order, as seen in the LDL phase) is significantly decreased.

Geske et al.\cite{Julian16} reported a similar $f(q_i)$ distribution structure (peak with a shoulder) for a higher temperature liquid silica ($T=5500$ K and ambient pressure), recognized as the fragile liquid. We note that, under the same temperature and pressure, the current SiO$_2$ HDL-LDL interface may hold a unique feature of microscopically  displaying the fragile-to-strong transition over spatial variation.\par

\begin{figure}[!htb]
	\centering
	\includegraphics[width=0.45\textwidth]{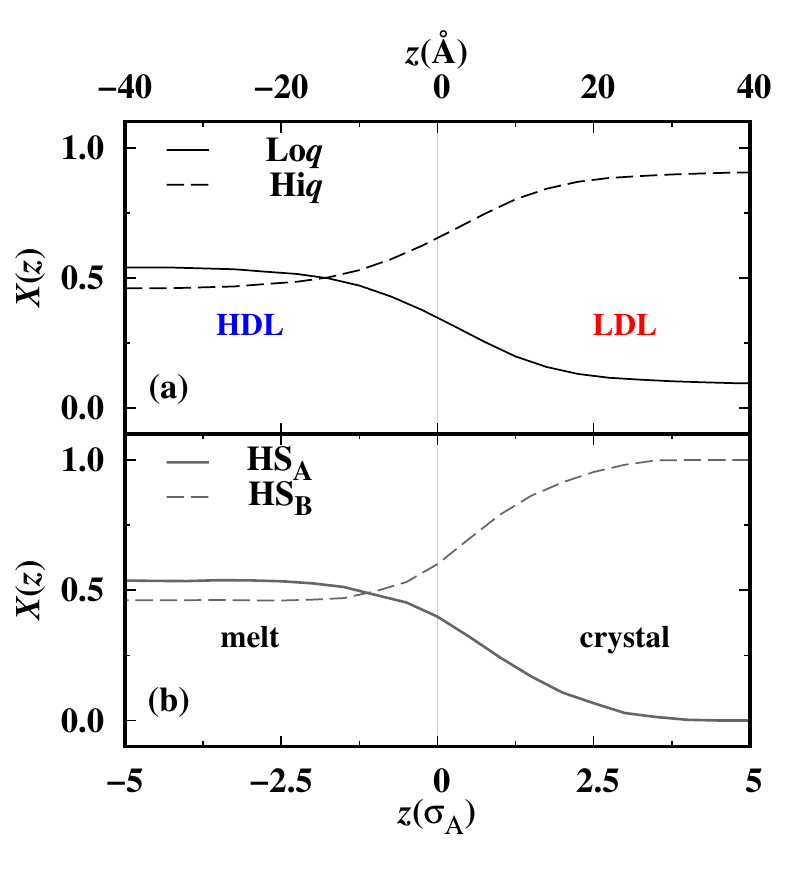}	
\caption{The silica HDL-LDL interface is analogous to the hard-sphere binary alloy melt-crystal interface. (a) The coarse-scaled concentration profiles across the HDL-LDL interfaces, the solid line and dashed lines are for the ``Lo'' atoms ($q_i<0.55$) and ``Hi$q$'' atoms ($q_i>0.55$), respectively. (b) The coarse-scaled concentration profiles across the fcc crystal-melt interface of a model hard-sphere binary AB alloy\cite{Laird02}. The solid line and dashed line are for the smaller A and larger B atoms, respectively.\label{4}}
\end{figure}

\begin{figure}[!htb]
	\centering
	\includegraphics[width=0.45\textwidth]{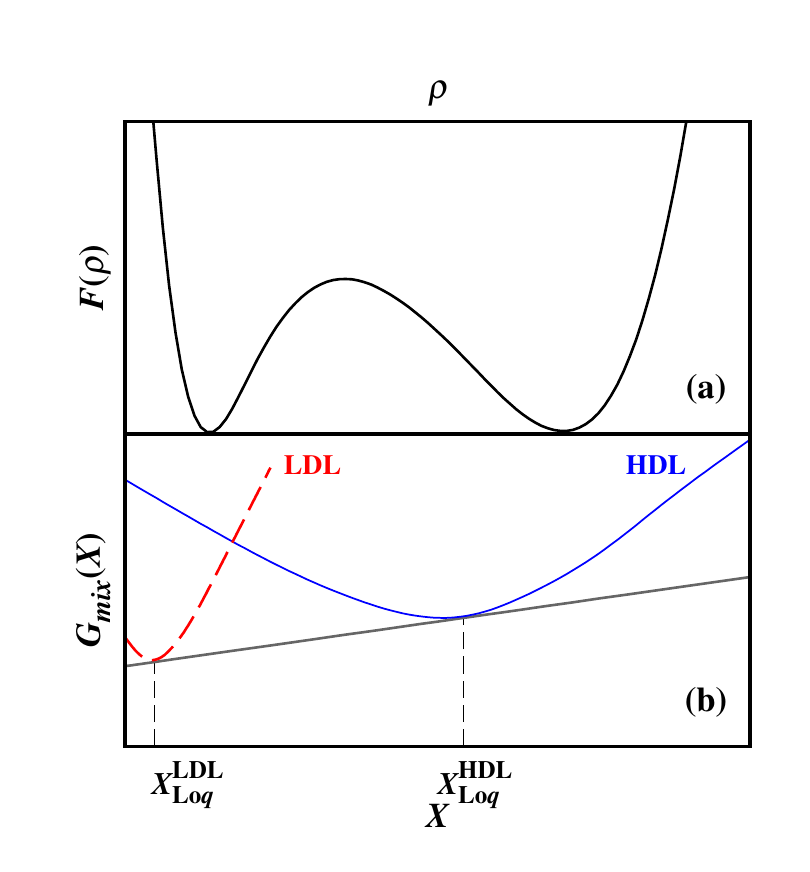}
\caption{(a) Schematic diagram of the commonly used free energy calculation methodology for searching the HDL-LDL phase coexistence conditions over order parameter space. (b) Current interpretation of the HDL-LDL coexistence draw on the analogy in FIG.\ref{4} as well as the traditional thermodynamics of alloys and phase equilibria, in terms of the concentration dependence of the mixing Gibbs free energies.\label{5}}
\end{figure}

Based on $q_i$ values, we separate the atoms into two categories, i.e., ``Hi$q$'' atoms for $q_i>0.55$, and ``Lo$q$'' atoms for $q_i<0.55$, and calculate the concentration profiles of each category, see in FIG.\ref{4}(a). In HDL region, ``Lo$q$'' and ``Hi$q$'' atoms possess the fractions of 54\% and 46\%, respectively. In LDL region, the concentration of ``Hi$q$'' atom approaches one ($\sim90\%$). Despite that this plot depicts the spatial transition from the strong tetrahedral order to the partially disrupted tetrahedral order, it is interesting to note that this plot bears a strong resemblance to that of the  binary hard-sphere interface between a single-component fcc crystal and the binary fluid mixture\cite{Laird02}, see in FIG.\ref{4}(b).

In order to understand the similarity between two systems shown in FIG.\ref{4}, we propose a following analogy. In the binary hard-sphere alloy system, the mixing is purely particle-packings. In FIG.\ref{4}(b), it is because of the significant size asymmetry (diameters differ by more than 85\%) that the binary alloy melt phase coexists with the crystal in which the small atom is immiscible. For the silica HDL-LDL interface system, we recognize the tetrahedron cluster connected in the network, as the ``larger building-block particle'' - analogous to the the hard-sphere with larger diameter. Whereas, the scattered (Si or O) atoms which do not belong to the tetrahedral network, are recognized as the ``smaller building-block particle'' - analogous to the hard-spheres with smaller diameter.

FIG.\ref{5}(b) depicts our interpretation of the HDL-LDL coexistence, in the frame work of binary alloy mixing free energetics, in addition to the well-known global free energy surface in ($\rho$ and/or $Q_6$) parameter space (panel a) \cite{Palmer14}. Therefore, the well-understood thermodynamics that predicts alloy phase diagram may be potentially transplanted to these complex network materials, giving potentially useful insight. For example, one might extend the our current analogy to the study of broader category of materials composing local network structures, i.e., medium-range ordering in metallic glass\cite{Ma11,Miao21}, gelation of colloidal particles\cite{Lu08}, and the emergence of the dynamic pre-ordering in the undercooled liquid water\cite{Matsumoto02,Fitzner19}, Si\cite{Rodrigo20} and Ga\cite{Niu20}.

\subsection{Dynamical Properties}\par

In this section, we characterize the diffusion coefficients, the structural relaxation time, and the atomic excitation indicator functions across the SiO$_{2}$ HDL-LDL interface, to depict the temporal and spatial variation in the dynamic properties of the interface.

\begin{figure}[!htb]
	\centering
	\includegraphics[width=0.43\textwidth]{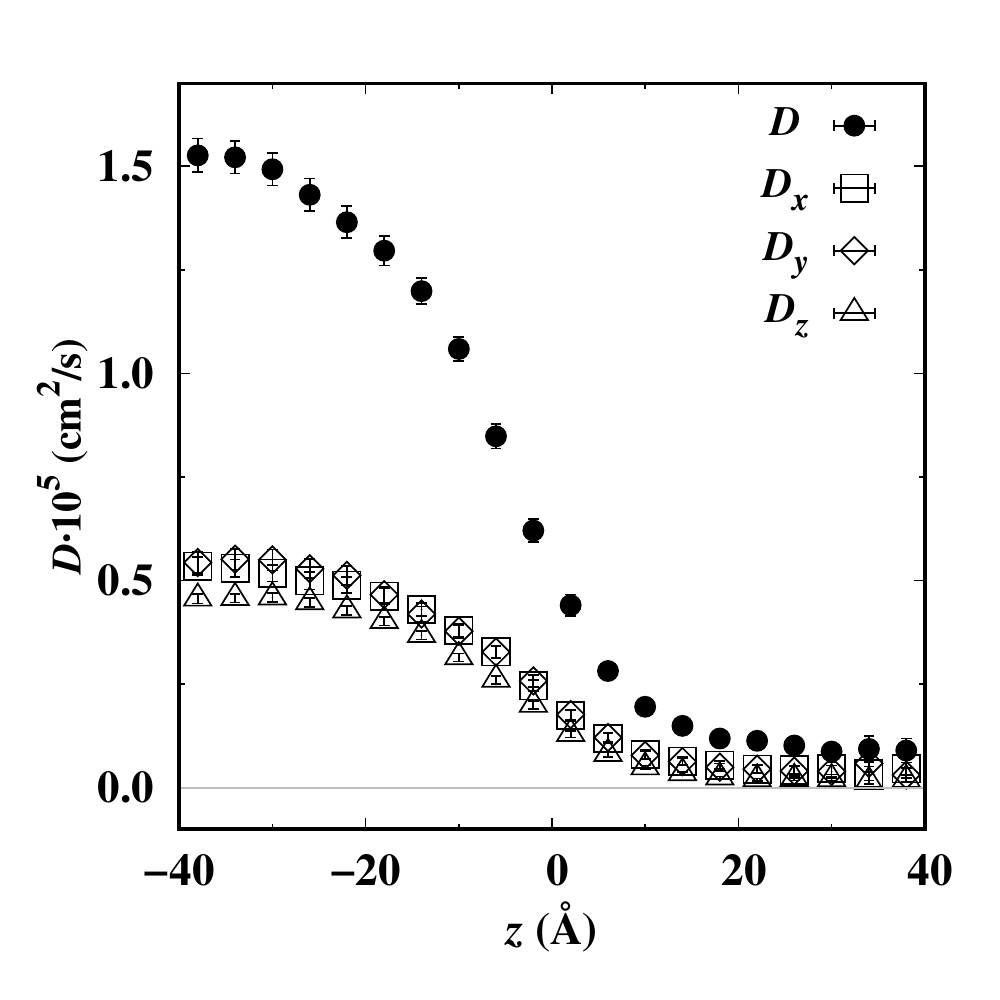}
	\caption{Coarse-grained diffusion coefficient profiles for the SiO$_2$ HDL-LDL interfaces. The total and three components of the diffusion coefficient are shown with different symbols. $z=0$ correspond to GDS, with $z<0$ for HDL and $z>0$ for LDL. The error bars represent 95$\%$ confidence levels. \label{6} }
\end{figure}

We calculate the MSD out to 10 ps, using Eq.(\ref{diff}), which is long enough to sample the mass transport in the diffusive regime, but short enough that the atoms do not travel no more than one bin spacing in $z$. FIG.\ref{6} shows the calculated diffusion coefficient profile for the HDL-LDL interface; in addition, the $x$, $y$ and $z$ components of the diffusion coefficient are also shown. The diffusion coefficient in the bulk SiO$_2$ HDL phase under (current $T=3100$K and $p=$0.4GPa) is $\sim1.5\times10^{-5}$ cm$^{2}$/s, which is a reasonable diffusion constant magnitude for typical liquids. Although the magnitude of $D\sim8.1\times10^{-7}$ cm$^{2}$/s for the bulk LDL phase is about 18 times smaller compared with that of the bulk HDL phase, it is as least 2 to 3 orders of magnitude greater than the (near-zero) diffusion coefficient in the silica crystal phase, suggesting that the LDL phase has liquid-like dynamics. At the same time, the $D(z)$ profile in the current homogeneous liquid-liquid interface system is qualitatively different to those seen in heterogeneous liquid-liquid interfaces (i.e. Al-Pb liquid-liquid interface\cite{Yang13}), in which interface layers show peak structures due to alloying of the two elemental species.

The $D(z)$ profile has the largest 10-90 width of 33.1\AA \ among all the profiles studied in this work. As the interface is traversed from HDL to LDL, the diffusion coefficient begins to decay at $z\sim-30$\AA \ until it converges to its bulk LDL value at $z>15$\AA. In contrast to the $Q(z)$ profile, the $D(z)$ profile is shifted towards the HDL phase relative to the GDS by 4.6\AA. Comparing all the midpoint positions listed in Table~\ref{tab1}, the mass transport, as measured by $D$, relaxes before all the rest of the properties as one travels across the interface from HDL to LDL.

Throughout the HDL-LDL interface, $D_x$ and $D_y$ are identical within the error bars, indicating the interface has x$\leftrightarrow$y symmetry. However, weak asymmetries between the normal and transverse components ($D_{x,y}=0.5(D_x+D_y)$ and $D_z$) on both sides of the HDL-LDL interface are seen, i.e., $D_{x,y}$ is greater than $D_z$ in both HDL and LDL phases, without overlapping in the error-bars. Through inspecting the center of mass for the LDL phase over a few ns during the equilibrium simulation, we identify that the whole LDL domain is undergoing a random walk along $x$ and $y$, which behaves similarly as a ``solid'' block undergoes Brownian motion in liquid, thus explaining the dynamical asymmetry, namely that $D_z < D_x$ and $D_y$ in both HDL and LDL sides

\begin{figure}[!htb]
	\centering	\includegraphics[width=0.45\textwidth]{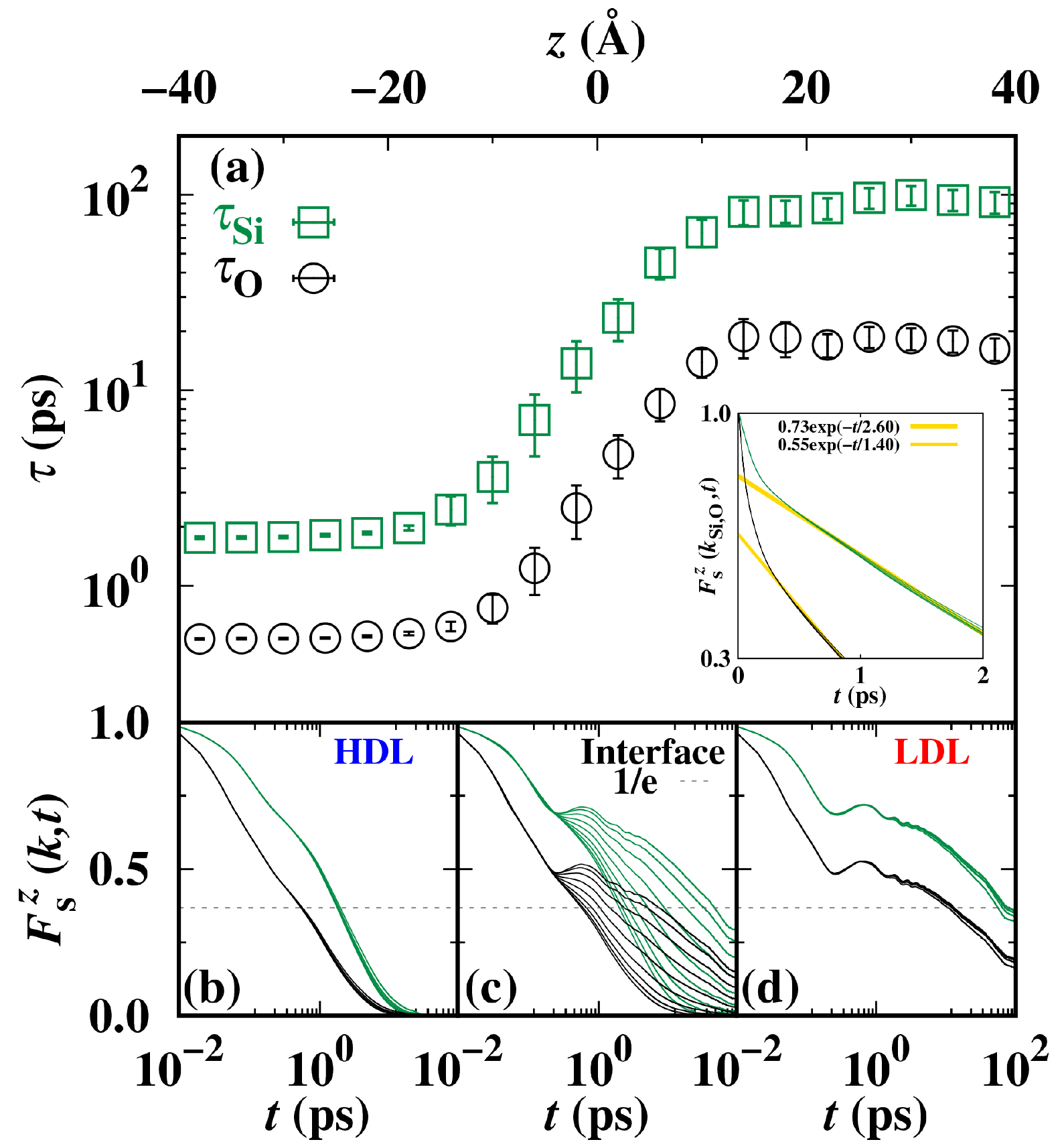}
	\caption{(a) Coarse-grained profiles for structure relaxation times across the SiO$_2$ HDL-LDL interface, the smaller circles represent $\tau_\mathrm{Si}$ for Si ions, the larger circles represent $\tau_\mathrm{O}$ for O ions. ISFs for the Si (green) and O (black) ions within different coarse-grained bins, $F^{z}_{s}(k_\mathrm{Si,O}, t)$, $k_\mathrm{Si}$=1.81 \AA$^{-1}$ and $k_\mathrm{O}$= 2.23 \AA$^{-1}$ for (b) the HDL region, (c) the interface region, (d) and the LDL region. The horizontal dotted line denotes a value of 1/e,  used for calculating the structure relaxation times in panel (a). The inset panel in (a) presents the Linear-Log plot of the same data shown in (b), in which the thick lines are fittings to exponential decay functions.\label{7}}
\end{figure}

In addition to the diffusion coefficient profile results, the heterogeneity of the spatial dynamics within the SiO$_2$ HDL-LDL interface can be characterized by calculating the structural relaxation times across the interface. FIG.\ref{7}(a) shows the coarse-grained profiles of the structure relaxation times for both silicon and oxygen atoms, $\tau_{\rm Si}(z)$ and $\tau_{\rm O}(z)$, respectively, across the HDL-LDL interface. The relaxation time in the bulk LDL phase is about two orders of magnitude larger than in the bulk HDL phases, because of the higher activation energy associated with breaking the tetrahedral bonds in the LDL phase\cite{Geske16}. As the interface is traversed from the LDL phase to HDL, both $\tau_{\rm Si}$ and $\tau_{\rm O}$ decrease as the HDL phase is approached. The increase in the structural relaxation rate can be attributed to the fact that it takes more energy to escape the deeper potential minimum, or break the tetrahedral bonds in the tetrahedrally bonded network in LDL (lower $\rho_e$) than in HDL, which is entropically stable.

In sharp contrast to the $D(z)$ and/or other profiles described earlier, the midpoint positions of the $\tau_{\rm Si}(z)$ and $\tau_{\rm O}(z)$ profiles lie on the LDL side, at $z$ positions (8.0\AA \ and 7.6\AA, respectively) significantly larger than those of the rest of the interfacial profiles listed in Tab.\ref{tab1}. Furthermore, the 10-90 widths of the two structure relaxation time profiles are less than half of that of the $D(z)$ profile. Therefore, comparing with mass transport across the HDL-LDL interface, the structural relaxation dynamics across the interface decays more sharply and is more localized. Note that, the decay region for both $\tau_{\rm Si}$ and $\tau_{\rm O}$ overlap significantly with the region where the coordination number ratio $n_\mathrm{Si}/n_\mathrm{O}$ reaches its highest value of 4.6.

Panel (b), (c), and (d) in FIG.\ref{7} display the calculated ISFs for the coarse-grained bins in the bulk HDL region, interface region, and the bulk LDL region, respectively. The ISF in bulk HDL is composed of a short-time non-exponential decay stage due to the vibrational dynamics, and a following exponential decay stage due to structural relaxation - see the inset panel in FIG.\ref{7}. The ISF (interfacial bins) extends over 1-2 orders of magnitude and the structural relaxation dynamics strongly slowdown as the HDL-LDL interface is traversed toward the LDL phase, developing in an intermediate plateau regime at the crossover between the short-time ballistic and the diffusive exponential decay stages. It is well-known that the intermediate plateau regime in the ISF is the consequence of the cage effect\cite{Gotze92}. Base on the previous knowledge of the temperature dependent ISF and $\tau$\cite{Geske17}, here the fragile-to-strong transition over spatial variation can be again recognized by the shape-evolution of the ISF as well as the increment of $\tau$ from HDL to LDL, consistent with the finding obtained from above $f(q_i,z)$ results.

Through labelling the Si atoms with the calculated excitation indicator function (Eq.\ref{h}), i.e., the left side of the FIG.\ref{8}, the mixture of the relatively immobile atoms (gray, $h_{i}(t)=0$) and the mobile atoms (white, $h_{i}(t)=1$) show significant spatial dynamical heterogeneity. All Si atoms in LDL have slow dynamics, which is consistent with the $D(z)$ and $\tau(z)$ results described above. In contrast to the LDL phase, about half of the Si atoms in HDL are highly mobile, while the remainder of the HDL atoms are not associated with excitation dynamics, it seems that these atoms in HDL interconnected with each other to form a network, and inherit the immobility from the LDL phase spatially through the interface.

\section{Summary}
 
We have presented a methodology for the calculation of structural, thermodynamic and dynamics properties for chemically homogeneous liquid-liquid interfaces. The methodology makes use of equilibrium molecular-dynamics (MD) simulations to characterize the spatial distribution profiles of various fundamental properties for the HDL-LDL coexistence interface of the LLPT system. This methodology was applied to the SiO$_2$ HDL-LDL interface modeled by modified WAC potential, at 3100 K and 0.4 GPa. Some of the principal results of this study are as follows:

\begin{itemize}
\item The interfacial profiles of different thermodynamics, structural and dynamics properties are seen to relax over varying length scales and are centered around at different positions relative to the Gibbs Dividing Surface.  As summarized in Tab.\ref{tab1}, the diffusion coefficient profile has the largest interfacial width whereas the structure relaxation time profile has the smallest width. The diffusion coefficients relax before the density, energy, the coordination numbers, and the structure relaxation times, as the interface is traversed from HDL to LDL. Overall, the HDL-LDL interfacial region is broader than any individual profiles.

\item We observe that the silica HDL-LDL interface displays a spatial fragile-to-strong transition through the interface by examining the results of both the probability distribution functions of the tetrahedral order parameter and the intermediate scattering functions over the coarse-grained bins across the interface. The decays of the structure relaxation dynamics are completed mostly on the LDL side towards the GDS of the interface, localized on a length scale of about 20\AA \ where the SiO$_2$ demonstrate remarkably mixing ability (of HDL and LDL).

\item Our analysis demonstrated that the interfacial stress profile $S(z)$, which usually exhibits significant structure in the interfacial region of liquid-liquid interfaces, was relatively flat indicating that the total excess interfacial stress, and thus the interfacial free energy is small. 

\item We categorized two types of ``building-block units'', on the basis of the tetrahedral order parameter, that make up both phases of the silica using an alloying mixture perspective, yielding an equilibrium concentration distribution that bears a striking resemblance to that of a model hard-sphere binary alloy system. With such conceptual analogy being made, we interpret of the LLPT phase coexistence in the framework of the traditional thermodynamics of alloys and phase equilibria.

\item Our calculations of dynamical properties reveal three kinds of temporal and spatial dynamical heterogeneities hybridizing within the silica HDL-LDL interface, i) temporal dynamics heterogeneities (stretched exponential ISF function) in LDL phase, ii) spatial dynamics heterogeneity across the HDL-LDL interface, and iii) mixture of high and low mobility atoms within the HDL phase.
%BBL09: What is a "temporal dynamical heterogeneity"? I think the terminology should be clarified here.
%Yang09: I have added "(or the stretched exponential ISF function)" to clarify the terminology. Is this OK?

%BBL10: Not sure what the final sentence above means.
% Yang10: I have deleted "The distinction between the structure-dynamics feature of the current liquid silica systems and those of the gelation system were discussed in detail."

\end{itemize}

The methodology proposed here should be applicable to more equilibrium atomic or molecular HDL-LDL interfaces, and potentially extendable to the exploration non-equilibrium LLPT systems.

\begin{acknowledgments}
We warmly thank Jun Ding and Xian-Qi Xu for helpful discussion. YY acknowledges the Chinese National Science Foundation (Grant No. 11874147), the Natural Science Foundation of Chongqing, China (Grant No. cstc2021jcyj-msxmX1144), Open Project of State Key Laboratory of Advanced Special Steel, Shanghai Key Laboratory of Advanced Ferrometallurgy, Shanghai University (SKLASS 2021-10), the Science and Technology Commission of Shanghai Municipality (No. 19DZ2270200, 20511107700) and the State Key Laboratory of Solidification Processing in NWPU (Grant No. SKLSP202105).
\end{acknowledgments}

\bibliography{ref}

\end{document}